# Multi-year Study of Environmental Stability of $Ti_3C_2T_x$ MXene Films

Asaph Lee, Mikhail Shekhirev, Mark Anayee, Yury Gogotsi*

*A.J. Drexel Nanomaterials Institute, and Department of Materials Science and Engineering, Drexel University, Philadelphia, Pennsylvania 19104, United States*

*Corresponding author: gogotsi@drexel.edu

**Abstract**

MXenes are a family of two-dimensional (2D) carbides and nitrides that display extraordinary electrical, optical, chemical, and electrochemical properties. There is a perception that MXenes are unstable and degrade quickly, limiting potential applications and requiring specific storage conditions to last for a long time. This was true for delaminated MXenes flakes in dilute dispersions prepared from defective precursors when MXene research was in its infancy. Since then, significant developments in MXene synthesis, processing, and understanding of its chemistry led to dramatic increases in environmental stability. Herein, we analyze $Ti_3C_2T_x$ free-standing films aged from 4 to 9 years through structural and morphological characterization along with electrical conductivity measurements to reveal the effect, or lack thereof, of prolonged storage under ambient conditions. Further, we show that the decrease in electronic conductivity over time is largely caused by the uptake of water by the hydrophilic surface chemistry of MXenes, which can be easily removed and its effect reversed by vacuum annealing.

**Introduction**

MXenes, first reported in 2011, are a family of two-dimensional materials composed of alternating layers of transition metals and carbon or nitrogen.[1] A diverse class of materials, they are known to have applications in areas such as electromagnetic interference shielding,[2] electrochemical energy storage,[3] gas- and bio-sensing.[4] The first discovered and most widely studied MXene composition is titanium carbide, $Ti_3C_2T_x$, produced by selective etching of Al from $Ti_3AlC_2$ MAX phase precursor.[5] Since 2011, the synthesis of $Ti_3C_2T_x$ has been extensively studied and optimized, leading to vastly improved physical and chemical properties including stability of colloidal dispersions of MXene flakes in water surpassing a year, and electronic conductivity of free-standing MXene films over 20,000 S cm$^{-1}$.[6] Despite this, there is still a perception that MXenes degrade quickly, on the order of weeks or even days.[7-9] While this is true for certain MXene compositions and at the early stages of MXene research, numerous strategies have since been developed to produce MXene dispersions and films that can be stable for months or years, and improvements in synthesis and processing of all MXene compositions are continually being made to reach the levels of $Ti_3C_2T_x$ MXene.[10]

Oxidation is the primary cause of MXene degradation over time.[11] Multiple factors affect oxidation rate, including MXene composition, form (powder, film, solution, etc.), and storage conditions.[12] Exposure to water may accelerate oxidation and also lead to hydrolysis of MXene. Through experimentation, it has been shown that when $Ti_3C_2T_x$ is in contact with water for a prolonged period, hydrolysis of the MXene layers assisted by the oxidizing effect of oxygen present in the water and air can facilitate $Ti_3C_2T_x$ conversion to titanium dioxide ($TiO_2$).[13,14] Oxidation can be quantified through conductivity measurements, as large conductivity drops have been shown by degradation of $Ti_3C_2T_x$ in air and when incorporated in solids such as polymers.[15] However, if the MXene is prepared in certain ways, degradation can be mitigated. Since as early as 2016, it has been known that drying at elevated temperatures can remove interlayer water from MXene films,[16] and that the removal of water can directly contribute to the enhancement of their conductivity[17] and environmental stability.[18] Vacuum drying can bypass potential oxidation due to the elevated temperature, but even drying $Ti_3C_2T_x$ in an oxygen-filled environment, such as ambient air, between 100°C and 245°C can improve conductivity by removing adsorbed and intercalated water and hydroxyl terminations without facilitating a decrease in its extinction coefficient or formation of $TiO_2$, which typically occurs above 450°C.[19]

One area of research that has developed rapidly in the past 5 years is the synthesis of the MAX phase precursors (for $Ti_3C_2T_x$, the MAX would be $Ti_3AlC_2$).[20,21] The properties of the MXene can vary significantly depending on the type of MAX used: for example, it was shown that MAX produced in three different ways (from graphite, carbon lampblack, and TiC) resulted in $Ti_3C_2T_x$ samples that varied in both conductivity and chemical stability.[20] In addition, the presence of aluminum exceeding the 3:1:2 stoichiometric ratios of Ti:Al:C during synthesis of the $Ti_3AlC_2$ MAX phase precursor resulted in superior crystallinity and carbon stoichiometry of the $Ti_3AlC_2$.[21] MXenes produced with this MAX displayed improved stability, and freestanding films and aqueous suspensions resisted oxidation for over 10 months.[21]

Another major change is to the etching method of the MXene itself. In 2018, a combination of lithium fluoride and hydrochloric acid,[22] significantly milder than the previously established etchant at the time, 50 wt.% pure hydrofluoric acid became the primary etchant used.[23] $Ti_3C_2T_x$ flakes synthesized in LiF/HCl (minimally invasive layer delamination or MILD method) demonstrated larger sizes as well as a lower concentration of point defects.[22,23] Three etchants with 5 wt.% HF were tested: pure HF, HF/HCl, and $HF/H_2SO_4$.[23] These etchants yielded various surface terminations, which in turn affected interlayer spacing and the amount of water present.[23] MXene etched in $HF/H_2SO_4$ had the cleanest surface with little to no sulfur, demonstrating the highest conductivity and the lowest amount of water.[23] However, the yield was dwarfed by HF/HCl, and so the currently established etchant for $Ti_3C_2T_x$ is a 6:3:1 ratio of HCl to deionized water to HF.[21] This 5 wt.% HF, HF/HCl solution combined with advances in MAX phase synthesis yielded stable $Ti_3C_2T_x$ films that exhibited conductivity exceeding 20,000 S cm$^{-1}$.[21]

At A.J. Drexel Nanomaterials Institute, there is convenient access to films made from the initial stage of MXene research in 2013, when the first delamination procedure was developed,[24] to the present, which grants a unique opportunity to study MXene films aged over many years. Over the years, we developed techniques to characterize the quality of MXene films.[25] In this study, the conductivity of $Ti_3C_2T_x$ MXene films synthesized from 2014 onward was measured and characterization of the samples that exhibited extremely high and low conductivity was obtained. Through X-ray diffraction and scanning electron microscopy, it was demonstrated that vacuum drying at the time of synthesis can slow degradation by removing interlayer water from the film, and drying even after 5-10 years have passed can still recover a significant amount of lost conductivity. Lastly, it was shown how synthesis and delamination methods to prepare the MXene films affect their conductivity and environmental stability over time.

**Experimental Methods**

*Film fabrication and collection.* 65 freestanding films of $Ti_3C_2T_x$ MXene produced between 2014 and 2021 were collected for analysis. Films were produced through two deposition techniques: (1) Vacuum-assisted filtration from colloidal dispersion of MXene in water through porous membranes (typically polyvinylidene difluoride (PVDF) with 450 nm pores);[26] (2) Spray coating from colloidal dispersion of MXene in water using a compressed air-backed spray gun onto Celgard membranes.[27]

*Conductivity measurement of $Ti_3C_2T_x$ films.* To acquire the conductivity of these films, the sheet resistance was measured using a Jandel four-point probe, and the thickness of the films was obtained using a Mitutoyo micrometer. In both cases, the measurements were taken at four separate points on the films and averaged.

*Drying of $Ti_3C_2T_x$ films* was done at 200°C for 24 hours in an Across International vacuum oven at less than 0.5 torr. After drying, the films were stored under a vacuum for 6 days before the XRD measurements were made.

*X-ray diffraction analysis* was performed using a Rigaku SmartLab diffractometer from 3° to 25° 2θ with a step size of 0.01°. The incident X-ray wavelength was 1.5406 Å. For the pre-drying data, the scan rate was 1° per minute, and for the post-drying data, the scan rate was 2° per minute.

*Scanning electron microscopy* was performed using a Zeiss Supra 50VP SEM. Samples were mounted on aluminum stubs using adhesive copper tape. Images were collected with an InLens detector at a working distance of 8-9 nm, a vacuum level of $1.5*10^{-5}$ Torr, and an accelerating voltage of 5.0 kV.

**Results**

Fig.1 summarizes conductivity measurements for the 65 films selected for this study. The wide range of conductivities of the measured $Ti_3C_2T_x$ films are indicative of a variety of synthesis methods, post-synthesis treatments, and storage conditions, due to varying purposes and applications. Many of them resulted from the search for optimal synthesis conditions and had properties far from the best. Films with the best properties were typically used for various research purposes, from destructive characterization to device fabrication, and not left intact. Thus, conductivity measurements and stability conclusions lean conservative, as there were likely samples from the same time period that would perform better than the 65 selected film samples. Three films of note are labeled in Fig. 1a; of these, film "A" had the highest measured conductivity of all measured sample films. With a conductivity of 9,750 S cm$^{-1}$, it compares favorably with the highest measured conductivities reported in the literature at the time of synthesis. Please note that literature reports measured the conductivity of freshly prepared films, while for the film "A" it is after 5 years of storage. This film was marked with its measured sheet resistance and thickness, which allows for a direct comparison of its current measured conductivity to conductivity at the time of synthesis. Film A was dried at the time of synthesis at 200°C for 24 hours. In contrast, film "B", part of the same batch as A, was prepared similarly to A but was kept at room temperature after synthesis. Lastly, film "C" demonstrated the lowest conductivity of all films, at 32 S cm$^{-1}$. This film was delaminated with tetramethylammonium hydroxide (TMAOH). Optically, all films appeared metallic and shiny, with no visible degradation, and maintained good flexibility. Photographs of exemplary films "A", "B", and "C" are shown in Fig. 1c.

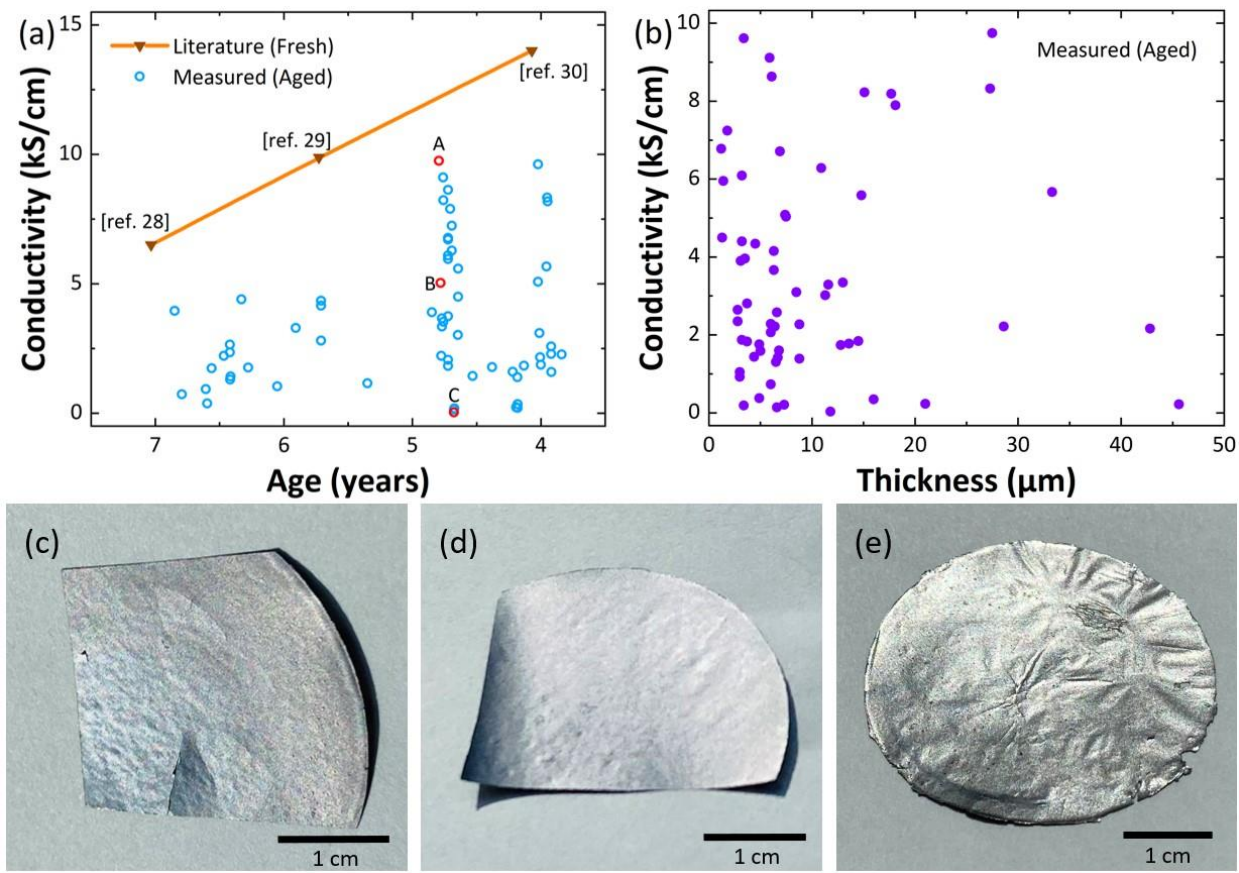

**Fig. 1** (a) Measured conductivity of aged $Ti_3C_2T_x$ films compared with maximum conductivity values from literature at the time of synthesis. (b) The conductivity of the aged films plotted vs. film thickness. Pictures of selected films from Fig. 1a; from left to right, film A (c), film B (d), and film C (e). All three films kept their metallic luster and did not show visual signs of oxidation.

There is a clear difference between the conductivities of measured films greater than and less than five years old (Fig. 1a). Films greater than five years old did not exceed 5,000 S cm$^{-1}$ (which is still an excellent value for the films aged 6-7 years), whereas the more recent films exhibited conductivities double that of the older films (Fig. S2). This steep increase in measured conductivity is likely due to the advances in MAX phase etching and MXene synthesis – the so known MILD method developed around that time, and the HF-HCl method used since 2021, which introduce less damage and produce higher quality MXene with a larger flake size and fewer point defects. In addition, the use of 5 wt.% HF etching solution[23] not only yields MXenes with superior conductivity but also greater stability, allowing them to retain much higher levels of conductivity over a prolonged time.

In Fig. 1b, conductivity was plotted vs. film thickness. Ideally, thickness should be inversely correlated with sheet resistance, resulting in an independent relationship between thickness and conductivity. However, as it is known that interlayer spacing can affect both conductivity and thickness, the correlation between conductivity and thickness was measured to determine whether it was necessary to adjust for thickness. The correlation value obtained was -0.018, close enough to zero that no adjustment was required. Thus, we can conclude that the film thickness did not affect the stability of MXene films and their properties after storage.

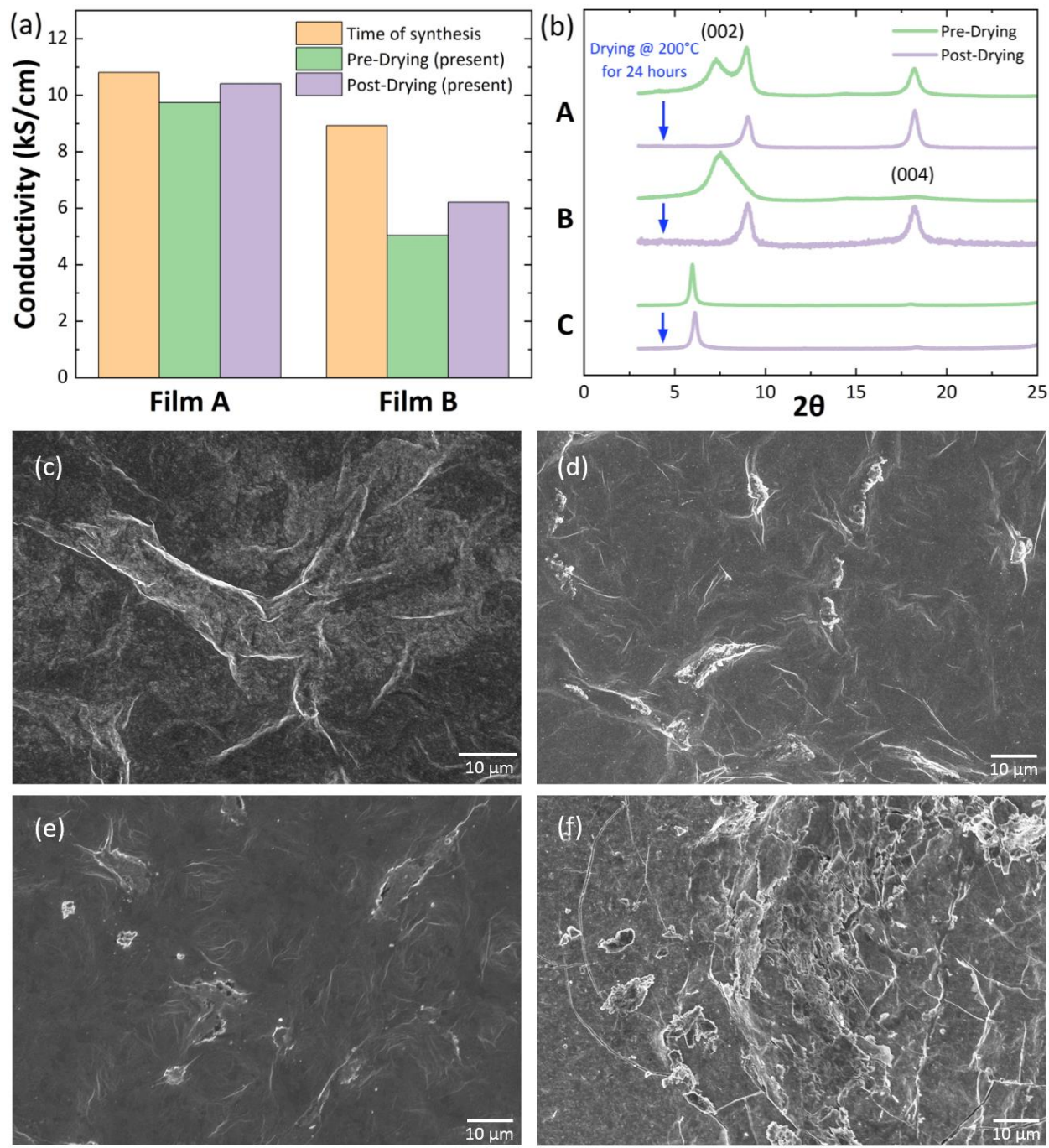

**Fig. 2** (a) Conductivity comparison between film A and film B at the time of synthesis, present-day before drying, and present-day after drying at 200°C for 24 hours under vacuum. (b) XRD of films A, B, and C both before and after drying at 200°C for 24 hours under vacuum. SEM images of film A (c), B (d), and C (e, f)

In addition to the information about the drying conditions, films A and B had conductivity data recorded at the time of synthesis, which allows for a comparison between the conductivities, shown in Fig. 2a (orange and green bars). Impressively, film A had a conductivity of 10,810 S cm$^{-1}$ when synthesized and a conductivity of 9,750 S cm$^{-1}$ when remeasured five years later - 90% of its original value. In contrast, film B retained roughly 50% of its original conductivity. This discrepancy appears to be solely due to the drying of sample A at an elevated temperature, which shows the importance of the removal of interlayer water. To measure interlayer spacing, X-ray diffraction analysis was performed on films A, B, and C, as shown in Fig. 2b. Of note, film A exhibited a double peak, at 7° and 9°, whereas film B had only a single peak at 7° 2θ. This indicates that film B's interlayer spacing is consistent, whereas certain areas of film A have a similar interlayer spacing to film B, but other areas have a smaller interlayer spacing. Combined with the conductivity data, water was likely present between the MXene sheets throughout film B, but only in certain areas in film A, which caused film B to lose conductivity to a greater degree than film A, presumably due to more extensive oxidation caused by the greater amount of interlayer water.

To confirm this, both films A and B were dried in a vacuum at 200°C for 24 hours and then recharacterized. The drying conditions were identical to the previous treatment of film A. Fig. 2a shows that after drying, both films recovered some of the conductivity lost over time. In Fig. 2b, the peaks in the XRD plots of the two films post-drying are virtually identical, and match peaks in film A pre-drying. This shows that the recovery of conductivity is likely caused by the removal of interlayer water via drying under vacuum for both films and the decrease of the interlayer spacing to 4.9 Å. Conductivity was not restored completely, likely because of oxidation over time. However, film A shows that drying both at the time of synthesis and after five years can result in 95% conductivity retention over that span. In addition, the XRD of film A pre-drying shows that there was still interlayer water after drying at the time of synthesis, but if dried to the point where water is completely removed right after synthesis, even better stability than what film A demonstrated could be attained.

Films A and B were both etched with hydrofluoric acid and sulfuric acid, and they exhibited superior conductivity to other films etched with hydrofluoric acid and hydrochloric acid. This method of using sulfuric acid produces clean surfaces with few surface terminations due to the removal of residual aluminum compounds. Hydrochloric acid, on the other hand, produces some chlorine terminations which increase interlayer spacing and may hydrolyze when water is imbibed due to the reactivity of the Ti-Cl bond.[23] It has been known that HF/H$_2$SO$_4$-etched Ti$_3$C$_2$T$_x$ exhibits superior conductivity, but this work shows that the difference in conductivity is preserved for five-year-old films. The advantage of HF/HCl is that the yield is high, currently ~75 wt.%,[21] and the current methods using HF/HCl can produce Ti$_3$C$_2$T$_x$ with a conductivity of 20,000 S cm$^{-1}$,[21] which is desired for most applications. However, if Ti$_3$C$_2$T$_x$ with extreme conductivity and stability is desired, etching with HF/H$_2$SO$_4$ or at least washing with H$_2$SO$_4$ after synthesis could be an option.

Scanning electron microscopy analysis was conducted on films A, B, and C. The SEM images of all films show a smooth surface throughout most, if not all, of the area; however, in film C, some areas showed a rough, disjointed surface with cracks and peeled-off layers. The method of delamination of film C is the likely cause. It was delaminated with tetramethylammonium hydroxide (TMAOH), a large organic molecule exceptionally adept at exfoliating multilayer MXene into its single-layer counterpart.[31] TMAOH is a versatile delaminating chemical, able to delaminate a wide variety of MXenes, but one downside is that the TMA$^+$ ions that intercalate the MXene layers to induce delamination are difficult to remove, thus resulting in a larger interlayer spacing than usual and easy water and air access to MXene surfaces. XRD of film C in Fig. 1b, with a single (002) peak at ~6°, showed an interlayer spacing of 7.4 Å, larger than either film A or film B. In addition, the peak didn't return to the position typical for the water-free film. It shifted by only about 0.1° after drying and the conductivity of film C increased from 32 S cm$^{-1}$ to 77 S cm$^{-1}$, signifying a significant increase due to water loss. In the MXene V$_2$CT$_x$, the larger interlayer spacing was shown to have contributed to the relatively lower conductivity values achieved, with a maximum of 140 S cm$^{-1}$.[32] In aged MXene films, the decrease in conductivity caused by a large interlayer spacing can be compounded by the increase in interlayer water permeation and oxidation rate due to said large interlayer spacing. In addition, film C delaminated with TMAOH, was likely made of MXene etched with a 50 wt.% HF solution,[31] which would result in smaller flakes and poorer conductivity than the 5 wt.% HF solution used for films A and B.[22,23] As shown by the comparison of these films, the choice of etchant and delamination method has a large effect on their stability and conductivity.

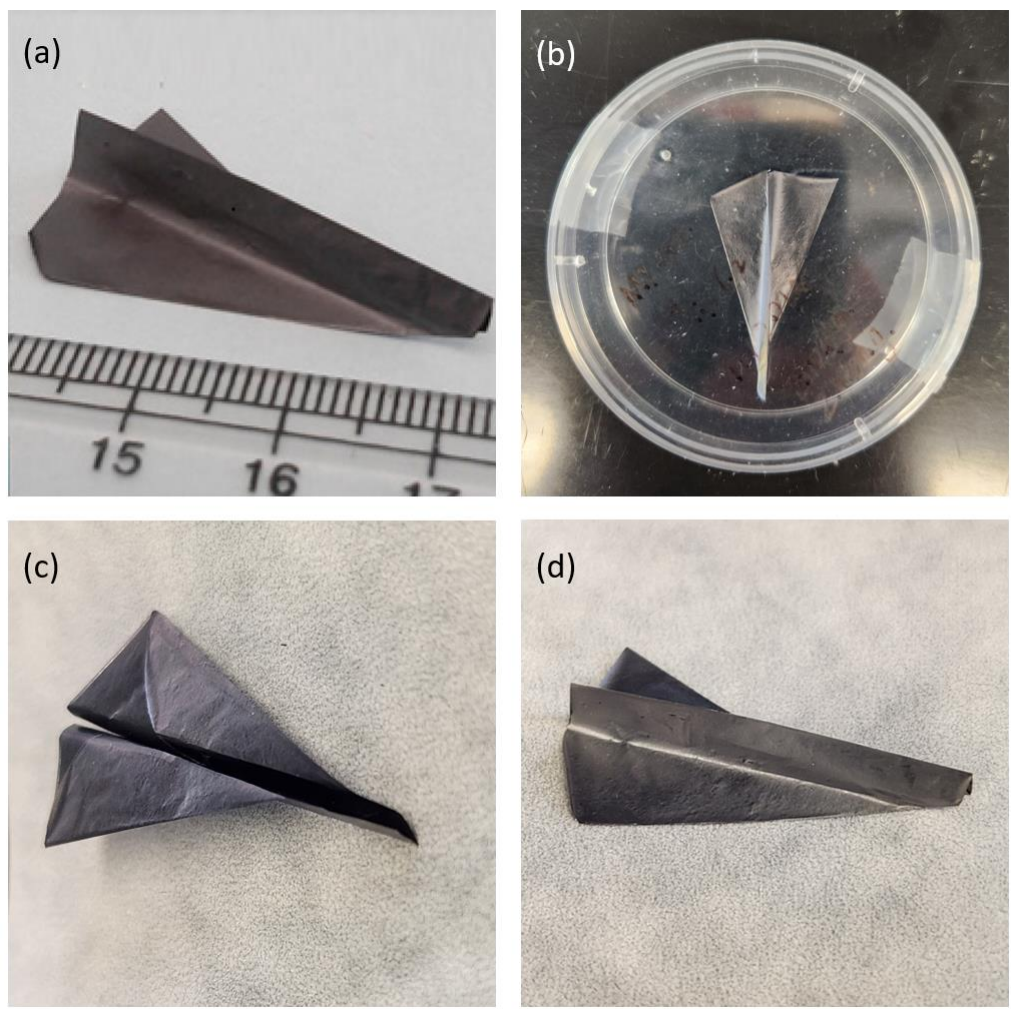

**Fig. 3** Photographs of a Ti$_3$C$_2$T$_x$ paper plane from 2013.[33] (a) At the date of synthesis.[33] (b) Stored in a petri dish since 2014. (c) Before drying. (d) After drying at 200°C for 24 hours to remove water.

One of the oldest published MXene foil samples avaiable at Drexel was a $Ti_3C_2T_x$ thin film fabricated in early 2014 and folded into a paper plane to demonstrate its flexibility and foldability.[33] At the time of synthesis, it exhibited a conductivity of 2,400 S cm$^{-1}$.[33] Remeasuring the conductivity about a decade later yields a value of 403 S cm$^{-1}$, a sixth of its original value, though Fig. 3b and 3c show its original appearance was maintained well and hardly any change could be visually observed. After drying at 200°C for 24 hours, the $Ti_3C_2T_x$ plane doubled its conductivity, reaching 849 S cm$^{-1}$, over a third of its original conductivity value. The role of removing water is shown here, as even conductivity loss over a decade could be significantly reversed. In addition, the sample itself displayed stability, as its shape or appearance didn't change, and it retained metallic conductivity over a span of a decade years. Take into account that was the earliest stage of MXene research when the quality of the delaminated $Ti_3C_2T_x$ was nowhere near the one we have today. Since then, there have been many developments that have served to increase the stability of MXenes. We expect that stoichiometric $Ti_3C_2T_x$ that showed no degradation for 2 months as single-layer flakes on a substrate[34] or over 10 months in dilute solution[21] will have a lifetime in the ambient environment well over a decade when processed into a film.

**Conclusions**

This work focuses on the characterization of $Ti_3C_2T_x$ MXene films produced over the past 10 years to demonstrate their stability and identify the factors that contribute to it. Electrical conductivity measurements were made to quantify the effect of degradation over time. The etching and delamination conditions have a great effect on the films' stability, and when prepared with the best practices, $Ti_3C_2T_x$ can retain virtually all of its conductivity after years of storage. Central to the stability of these films over time is control of the interlayer spacing, as a small interlayer spacing prevents water from entering and facilitating oxidation and degradation. On the other hand, the use of TMAOH or TBAOH, whose ions increase interlayer spacing and provide access to water and oxygen, leads to the degradation of the films over time. Using milder etchants that decrease the concentration of defects and yield larger flakes, compared to the 50 wt.% HF used in the first few years of MXene research, also prolongs the lifetime of MXene films.

$HF/H_2SO_4$-etched $Ti_3C_2T_x$ displayed remarkable stability with 90% of initial conductivity (~10,000 S cm$^{-1}$) after 5 years in the laboratory environment. This suggests that clean surfaces with no chlorine terminations or residual aluminum etching products lead to the highest environmental stability. Chlorine and fluorine surface terminations may hydrolyze over time if water can access the interlayer spacing, and create nucleation sites for oxide formation. A decrease of the interlayer spacing after drying at elevated temperature resulted in MXene films that retained high conductivity. This method can be used to produce highly conductive and environmentally stable MXene. Drying at 200ºC and higher temperatures is a simple and powerful tool to increase the stability of MXenes. It removes interlayer water and welds the MXene layers together, guarding against future degradation.

The fact that drying can restore the conductivity of aged MXene films suggests that the observed decrease in conductivity is largely due to water penetration between the hydrophilic MXene layer, not the degradation of MXene. When oxidation or hydrolysis products are formed between MXene sheets, the lattice spacing cannot be recovered by drying. We anticipate that polymer coatings or any packaging methods used in the semiconductor industry that protect the material from humidity are expected to greatly prolong the lifetime of MXene films.

## Acknowledgments

Samples were synthesized by Dr. Kanit Hantanasirisakul, Dr. Narendra Kurra, Dr. Kathleen Maleski, and other researchers at the A.J. Drexel Nanomaterials Institute. The $Ti_3C_2T_x$ paper plane was made by Dr. C. Evelyn Ren and stored in Prof. Yury Gogotsi's office for the past decade.

**Supporting Information**

**Multi-year Study of Environmental Stability of Ti$_3$C$_2$T$_x$ MXene Films**

Asaph Lee, Mikhail Shekhirev, Mark Anayee, Yury Gogotsi

*A. J. Drexel Nanomaterials Institute, and Department of Materials Science and Engineering, Drexel University, Philadelphia, Pennsylvania 19104, United States*

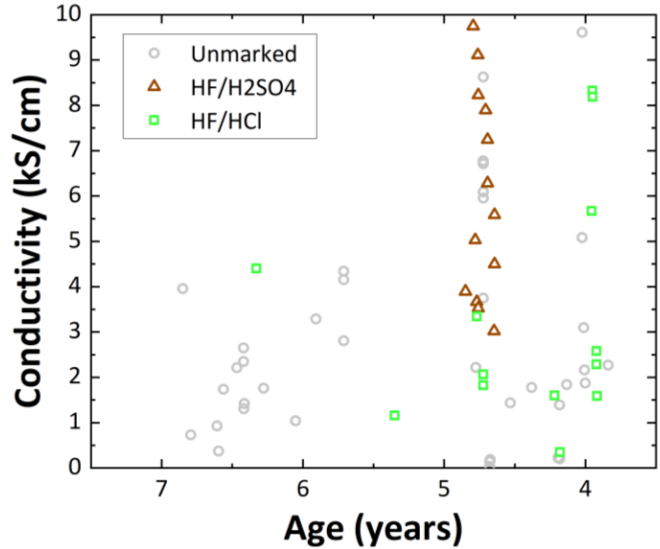

**Fig. S1** Conductivity difference between HF/H$_2$SO$_4$ and HF/HCl-etched Ti$_3$C$_2$T$_x$ due to differences in the surface chemistry.

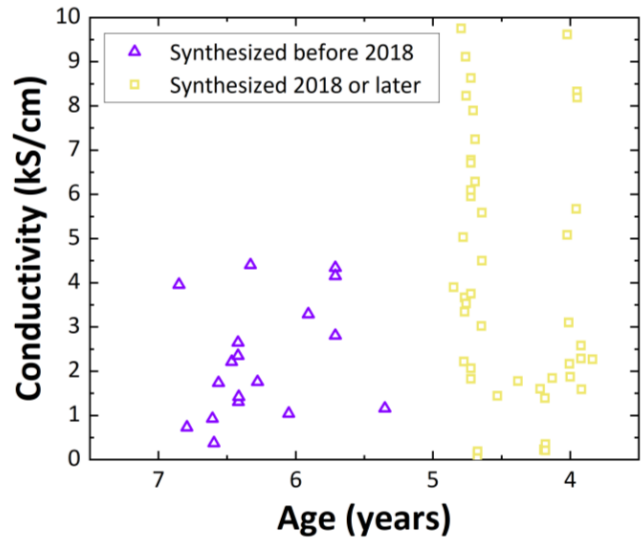

**Fig. S2** Conductivity difference between the MXene samples prepared before and after 2018 due to improvements in MAX phase synthesis and etching process.